\documentclass{cimento}

\usepackage{graphicx}
\usepackage{epstopdf}

\title{Non-local probes for a relaxing non-Abelian plasma}
\author{L.~Bellantuono}
\instlist{\inst{} Universit\`a di Bari e INFN, Sezione di Bari - Bari, Italy}

\PACSes{\PACSit{12.38.Mh}{Quark-gluon plasma}
}

\begin{document}

\maketitle

\begin{abstract}
The thermalization of a strongly coupled plasma is examined in the holographic framework through non-local observables: the equal-time two-point correlation function of a large dimension boundary operator, and Wilson loops of different shapes. 
The evolution of the probes from an initial far-from-equilibrium state to a hydrodynamic regime is found to depend on their size. A hierarchy among the thermalization times of the energy density, the pressures and the large size probes, is identified: the relaxation process is faster at short distances.  
\end{abstract}

\section{Boundary sourcing and local probes}
In recent experiments of relativistic heavy ion collisions realized at the Brookhaven RHIC and at the CERN LHC, the production of a deconfined and strongly coupled quark-gluon plasma (QGP) has been detected. The QGP expands anisotropically and cools, reaching the hydrodinamic regime after a time of order 1 fm/c. The relaxation towards equilibrium can be studied in the holographic approach, a framework inspired by the gauge/gravity correspondence. This conjecture relates a strongly coupled Super-Yang Mills gauge theory defined in a $4$-dimensional Minkowski space (boundary) with a dual classical gravity theory living in a $5$-dimensional anti-de Sitter space (bulk) times a compact manifold~\cite{ref:Maldacena}.
The evolution of the QGP initial state, anisotropic and far from equilibrium, can be examined by introducing an impulsive perturbation (quench) to the metric on the boundary, and then solving the Einstein equations in the bulk~\cite{ref:Chesler}. In our analysis, the external sources that distort the boundary metric are chosen to mimic processes in which a small number of collisions occur. As the quench becomes static, the system starts to relax towards the hydrodinamic regime.
The boundary coordinates are denoted as $x^{\mu}=\left(x^{0},x^{1},x^{2},x^{3}\right)$, with $x^{3}=x_{||}$ the collision direction. We assume that the geometry is invariant under boost transformations along this axis and under translations and rotations in the transverse plane $\textbf{x}_{\perp}=\left(x_{1},x_{2}\right)$. The Minkowski line element $ds_{4}^{2}=-d\tau^{2}+dx_{\perp}^{2}+\tau^{2}dy^{2}$, with $\tau$ the proper time and $y$ the spacetime rapidity $(x^{0}=\tau \mathrm{cosh} y$ and $x_{||}=\tau \mathrm{sinh} y )$, is perturbed through the quench $\gamma(\tau)$ as follows:
\begin{equation}\label{e.4d}
ds_{4}^{2}=-d\tau^{2}+e^{\gamma(\tau)}dx_{\perp}^{2}+\tau^{2}e^{-2\gamma(\tau)}dy^{2}.
\end{equation}
The stress-energy tensor on the boundary yields a local probe of the relaxation process and a measure of the residual anisotropy in the plasma. Its components, the energy density and the pressures longitudinal and transverse with respect to the collisional axis, have been computed for different quench profiles and compared to the viscous hydrodinamics behaviour. Regardless of the considered distortion, hydrodinamization of the stress energy-tensor is a two-step process: the energy density acquires the viscous form as soon as the quench is switched off, while pressure isotropy is restored with a time delay of order of $1$ fm/c, for an effective temperature at the end of the quench of about $500$ MeV~\cite{ref:Bellantuono2015}. 
Such description can be compared to the one deduced from the time evolution of non-local observables~\cite{ref:Balasubramanian}, as presented in the next section for a particular distortion profile~\cite{ref:Bellantuono2016}. 

\section{Non-local probes}
\begin{figure}
\centering
\includegraphics[width=0.9\textwidth]{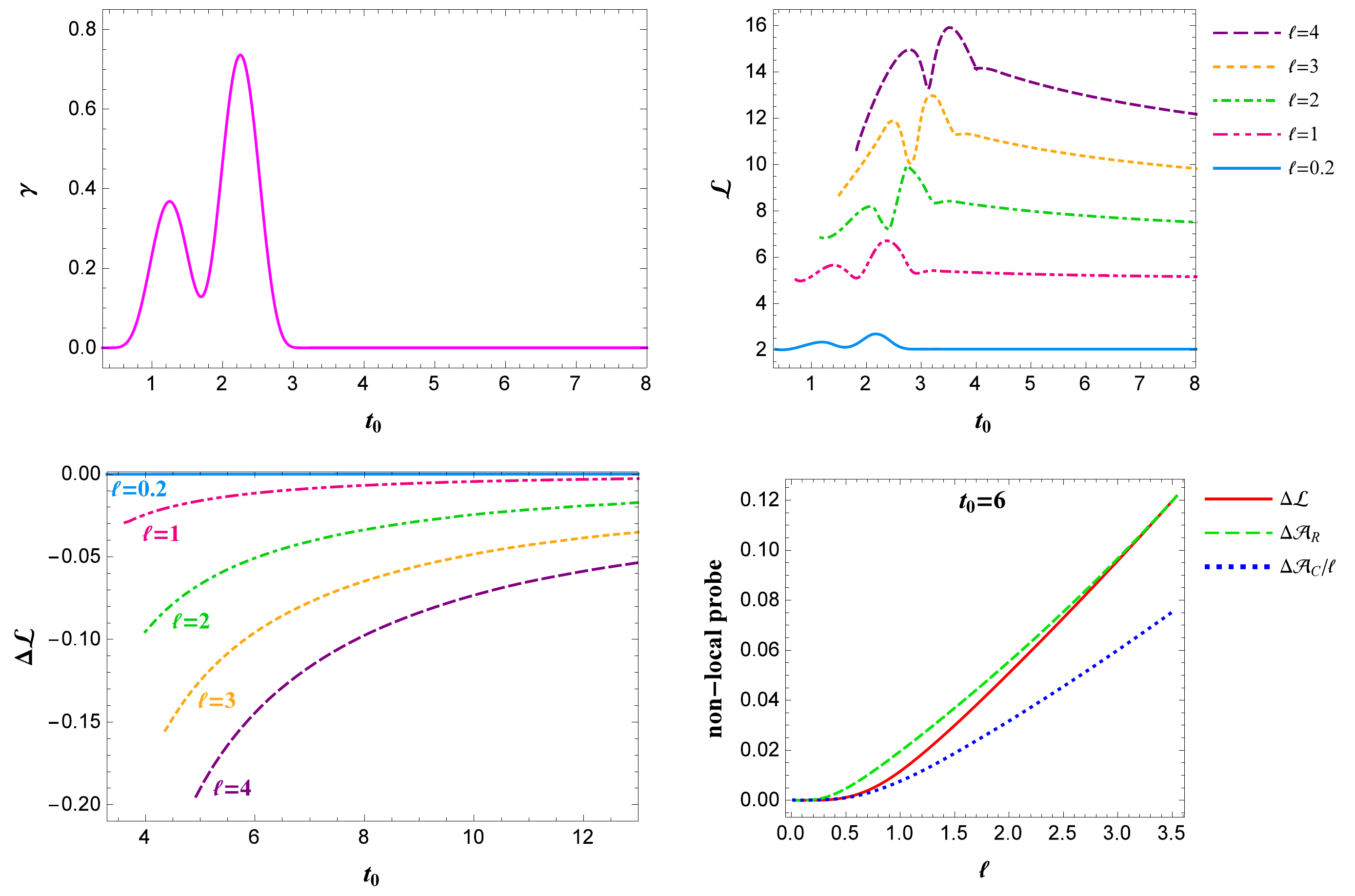}     
\caption{Upper left: profile of the quench $\gamma$. Upper right: geodesics length $\mathcal{L}$ for different vales of the size $\ell$. Lower left: difference $\Delta\mathcal{L}$ between the geodesics lengths of size $\ell$ in the quenched and in the hydrodynamic geometries. Lower right: non-local probes $\Delta\mathcal{L}$, $\Delta\mathcal{A}_{\mathrm{R}}$ and $\Delta\mathcal{A}_{\mathrm{C}}/\ell$ at the isotropization time $t_{0}=6$.}\label{fig}
\end{figure}

The relaxation process can be monitored through non-local probes, such as the two-point correlation function of boundary theory operators, and expectation values of Wilson loops defined on the boundary~\cite{ref:Bellantuono2016}. The system under investigation is driven out-of-equilibrium by boundary sourcing using a sequence of two nearly-overlapping pulses, as shown in fig.~\ref{fig} (top left). 

We consider the correlation function of a boundary scalar operator $\mathcal{O}$ with large conformal dimension $\Delta\gg 1$, between two equal-time points having a spatial separation $\ell$ in the transverse plane: $P=\left(t_{0},-\ell/2,x_{2},y\right)$ and $Q=\left(t_{0},\ell/2,x_{2},y\right)$. In the holographic picture, this observable can be approximated as $\langle \mathcal{O}(P)\mathcal{O}(Q) \rangle \simeq e^{-\mathcal{L}\left(t_{0},\ell\right)\Delta}$, where $\mathcal{L}\left(t_{0},\ell\right)$ is the length of the geodesic that connects the boundary points and extends in the bulk at fixed $\left(x_{2},y\right)$. Figure~\ref{fig} (top right) shows the evolution of $\mathcal{L}$ for several values of the distance $\ell$ between the points in the correlation function: the curves follow the distortion profile, with a time delay that increases with the size of the probe. In order to evaluate the degree of thermalization of the system, it is interesting to examine the difference $\Delta\mathcal{L}$ between the length $\mathcal{L}$ in the geometry under investigation and the same observable computed in a bulk metric reproducing the viscous hydrodynamic time dependence of the stress-energy tensor~\cite{ref:Bellantuono2016}. The evolution of $\Delta\mathcal{L}$ after the end of the quench is shown in fig.~\ref{fig} (bottom left) for different values of $\ell$. Each curve is plotted starting from the value of the physical time $\tilde{t}_{0}(\ell)$, at which it is no longer affected by the quench. 

An analogous approximation can be used to compute the expectation value of the Wilson loop along a closed contour $\mathcal{C}$ of linear size $\ell$ and fixed time $t_{0}$ on the boundary: $\langle W_{\mathcal{C}} \rangle \simeq e^{-\mathcal{A}\left(t_{0},\ell\right)}$, with $\mathcal{A}\left(t_{0},\ell\right)$ the area of the extremal surface bounded by $\mathcal{C}$ and plunging in the bulk at fixed $y$. Two shapes for $\mathcal{C}$ have been considered: a rectangular strip of finite width $\ell$ along the $x_{1}$ axis and infinite length along $x_{2}$, and a circular path of diameter $\ell$ in the transverse plane $\textbf{x}_{\perp}=\left(x_{1},x_{2}\right)$. The areas of the extremal surfaces bounded by these contours, $\mathcal{A}_{\mathrm{R}}$ and $\mathcal{A}_{\mathrm{C}}$ respectively, and the differences $\Delta\mathcal{A}_{\mathrm{R}}$ and $\Delta\mathcal{A}_{\mathrm{C}}$ between the observables in the quenched and in the hydrodynamic geometries, have been worked out in~\cite{ref:Bellantuono2016}. The behaviour of such quantities resembles the one in fig.~\ref{fig} (top right and bottom left) for the geodesics case.

The time at which each non-local probe reaches the viscous hydrodynamic regime depends on its size: for larger boundary separation between the geodesics extremes or among the contour points, the quantities $\mathcal{L}$, $\mathcal{A}_{\mathrm{R}}$ and $\mathcal{A}_{\mathrm{C}}$ take longer to equilibrate. The thermalization time can be identified with the half-life $t_{1/2}(\ell)$, defined as the value of $t_{0}$ at which $\Delta\mathcal{L}$, $\Delta\mathcal{A}_{\mathrm{R}}$ and $\Delta\mathcal{A}_{\mathrm{C}}$ are reduced by a half with respect to their values at time $\tilde{t}_{0}(\ell)$. The analysis reveals that $t_{1/2}(\ell)$ exceeds the thermalization time obtained from the stress-energy tensor for boundary separations $\ell\simeq1$, and increases linearly with $\ell$ for larger probes. A hierarchy among the relaxation times of the energy density, pressures and large probes emerges, indicating that thermalization is faster at UV scales.
Although the three non-local observables have the same qualitative behaviour with $t_{0}$ and $\ell$, their half-lives and thermalization length scales are quantitatively different. For a comparison, we show in fig.~\ref{fig} (bottom right) the $\ell$-dependence of the quantities $\Delta\mathcal{L}$, $\Delta\mathcal{A}_{\mathrm{R}}$ and $\Delta\mathcal{A}_{\mathrm{C}}/\ell$ at the physical time $t_{0}=6$ at which pressure isotropy is restored in the plasma~\cite{ref:Bellantuono2015}. It is possible to identify for each function a transition to a linear regime. The inflection point of the first derivative can be used to characterize the onset of the linear behaviour, and defines the length scale above which the probe is not thermalized at the chosen time~\cite{ref:Bellantuono2016}. The rectangular Wilson loop thermalizes slower than the other non-local probes.

\acknowledgments
I thank P. Colangelo, F. De Fazio, F. Giannuzzi and S. Nicotri for collaboration.

\end{document}